\documentclass[%
twocolumn,
amsmath,amssymb,
aps
]{revtex4}

\usepackage{graphicx}
\usepackage{dcolumn}
\usepackage{bm}
\usepackage{braket}
\usepackage{float}
\usepackage{mathtools}
\usepackage{hyperref}
\usepackage{subfig}
\usepackage{enumitem}   

\begin{document}
	
	
	\title{Interference-induced localization in quantum random walk on clean cyclic graph}
	
	\author{Jayanth Jayakumar}
	\author{Sreetama Das} \author{Aditi Sen(De)} \author{Ujjwal Sen}
	\affiliation{%
		Harish-Chandra Research Institute, HBNI, Chhatnag Road, Jhunsi, Allahabad 211 019, India
	}%

	
	\begin{abstract}
		We quantitatively differentiate between the spreads of discrete-time quantum and classical random walks on a cyclic graph. Due to the closed nature of any cyclic graph, there is additional ``collision"- like interference in the quantum random walk along with the usual interference in any such walk on any graph, closed or otherwise. We find that the quantum walker remains localized in comparison to the classical one, even in the absence of disorder, a phenomenon that is potentially attributable to the additional interference in the quantum case. This is to be contrasted with the situation on open graphs, where the quantum walker, being effectively denied the collision-like interference, garners a much higher spread than its classical counterpart. We use Shannon entropy of the position probability distribution to quantify spread of the walker in both quantum and classical cases. We find that for a given number of vertices on a cyclic graph, the entropy with respect to number of steps for the quantum walker saturates, on average, to a value lower than that for the corresponding classical one. We also analyze variations of the entropies with respect to system size, and look at the corresponding asymptotic growth rates.
	\end{abstract}
	
	\maketitle
	
	
	\section{\label{sec:level1}Introduction}
	
	Quantum interference is one of the key aspects of quantum mechanics that leads to a plethora of interesting phenomena, such as interference pattern in double-slit experiments and Anderson localization of electron wave packets \cite{ref1,ref2}. 
	It also leads to the qualitatively different behavior of 
	quantum random walks (QRWs), first introduced by Aharonov, Davidovich, and Zagury \cite{ref211} in 1993, from their classical counterparts - the 
	classical random walks (CRWs) \cite{ref3}. 
	Superposition over walker-coin states where the walker has pursued different paths as instructed by the different coin states, and the resulting entanglement \cite{ref212} between the coin and position degrees of freedom in a discrete-time QRW are at the root of its differences with a walker in the classical case. In particular, there is inherent interference between the left and the right propagating components in the dynamics of a QRW on a line, and the clockwise and anti-clockwise components on a circle (cyclic graph). Based on the chosen initial conditions or by varying the coin parameters, interference may affect the symmetry of the probability distribution, of the walker on both the line and the cyclic graph, in position space \cite{ref711, ref712, ref713, ref7}. QRWs have been shown to be useful in realizing quantum memory \cite{ref11}, in search algorithms \cite{ref16, ref17, ref18, ref91} 
	(see also \cite{ref19,ref21,ref20}), for simulating dynamics of physical systems \cite{ref22,ref23,ref24}, etc.
	Successful experimental implementation of QRWs 
	%
	have been reported in various physical systems, such as optical Galton board \cite{ref240}, nuclear magnetic resonance systems \cite{ref241, ref242}, atoms trapped in an optical lattice \cite{ref243}, photons \cite{ref244, ref245, ref246, ref247, ref248, ref249}, and trapped ions \cite{ref250, ref251}. Proposals for experiments include \cite{ref253, ref254, ref255}.  Introduction of disorder in the system, e.g. by randomizing the ``coin parameter" of a discrete-time quantum random walk \cite{ref8}, or by introducing static disorder in a continuous-time  quantum random walk on the glued trees graph \cite{ref9}, or by inserting other imperfections on graphs where the distances between the vertices vary slightly from edge to edge \cite{ref81} may lead to ``localization" of the wave packet in position space. See also \cite{ref82,ref83,ref84,ref85,ref86} in this regard. Notice that this is analogous to the phenomenon of localization in condensed matter physics \cite{ref10}.

	\par For a discrete-time quantum random walk on a line, the spread of the probability distribution, as quantified by the standard deviation, scales linearly with the number of steps of the walker, referred to as ``ballistic spread". The corresponding ``speed" of the walker is only square-root of the number of steps for a classical walker \cite{ref25,ref711}. It is possible that for quantum and classical random walks on a cyclic graph, the scaling behaviors with respect to steps or system-size may hide interesting information. The long-time properties of the time-averaged probability distributions in both the cases have already been studied \cite{ref271,ref272,ref273}. Quantum walks on a cyclic graph having the additional feature of ``one-step memory" was investigated in \cite{ref35}. By allowing the coin operation to change at every step according to a sequence or by random means, the associated probability distribution is seen to converge to a uniform distribution over the nodes of a cyclic graph \cite{ref37}. In another work, the periodicity of the evolution matrix of a Szegedy walk, a type of discrete-time quantum walk, on various types of finite graphs have been discussed \cite{ref38}. A special feature of QRW on a cyclic graph is that it ``mixes" almost quadratically faster than the corresponding classical case \cite{ref28}. QRWs on cyclic graphs have been implemented experimentally, and further proposals thereof have also been given. In an experimental implementation using an arrangement of linear optical elements, clockwise and anti-clockwise cyclic walks have been realized \cite{ref33}. In another experimental implementation, continuous-time quantum walks on cyclic graphs using quantum circuits have been realized \cite{ref331}. Travaglione \textit{et al.} have proposed a scheme to implement QRW on a line and on a circle in an ion trap quantum computer \cite{ref252}. Another proposal for experimental implementation of QRW on a cyclic graph, using a quantum quincunx, which may be realized with cavity quantum electrodynamics, is also known \cite{ref256}. QRWs of non-interacting and interacting electrons on a cyclic graph, with the graph being formed of semiconductor quantum dots, have been studied in \cite{ref34}.
	
	\par In this paper, we analyze the behaviors of quantum and classical random walks on  cyclic graphs.
We find that the
``collision"-like interference effects on the cyclic lattice, occurring due to the topology of the lattice, leads to a ``localization'' behavior of the wave packet in a Hadamard quantum walk. The localization behaviors, both with respect to number of steps and number of sites, are inferred by analyzing the patterns of Shannon entropies of the corresponding position probability distributions. This is in contrast to the well-studied localization induced by some form of disorder introduced in the system. We then show that the behavior is generic, in that it appears also 
for non-Hadamard walks, and that it is independent of the initial state.
Moreover, we find that the scaling exponents of the entropies with respect to system size, when the number of steps is sufficiently large,  are different for classical and quantum random walks. Furthermore, the ratio between these entropies again implies that the quantum system is localized with respect to the classical one. 
Finally, we introduce a distance measure using the $l^1$-norm, for measuring the closeness of two distributions, specifically, the time series data of Shannon entropies for a QRW and a CRW on cyclic graphs of increasing size. Using this measure, we quantify the difference in the spreads of QRW and CRW on cyclic  lattices of different size.

The paper 
is structured as follows. Section \ref{sec:levell1} introduces the general aspects of a CRW on a cyclic graph and its operational formalism. Section \ref{sec:levell4} introduces the broad facets of a QRW on a cyclic graph, the operational formalism, and describes 
the localization behavior with respect to steps 
in Hadamard and non-Hadamard quantum walks.
In section \ref{sec:levell5}, we study the behavior of CRW and QRW on cyclic graphs with respect to the system size.
Section \ref{sec:levell3} analyzes 
closeness of the two time series data distributions by inspecting the corresponding \(l^1\)-norm.
We present a summary in Section \ref{sec:level2}. 

	\section{\label{sec:levell1}Classical random walks}

	 \par We begin this section by giving a short introduction to CRW on the infinite line. We next introduce the cyclic graph and discuss about the features of the probability distribution of the walk after a large number of steps. We also briefly indicate the mathematical formalism that is used to deal with CRWs on the cyclic graph.
	
	\par The mathematical setting for random walks are graphs $G(V,E)$ with a vertex set $V$ and an edge set $E$. Classical random walks consist of a walker localized at a given vertex $v$ who moves by means of randomly choosing one of the two directed edges, for example, of an infinite line or a cyclic graph, with probabilities $p$ and $q$ $(p+q=1)$ at every step. We are restricting ourselves to situations where there are exactly two edges emanating from every vertex, and the graph is connected. This motion is dictated by the result of toss of a coin which could be unbiased or not. Let us however assume that the coin is unbiased, so that $p=q=1/2$. A natural question of interest is the following: What is the pattern of probability distribution of the walker over all the vertices after a given number of coin tosses? It turns out that for relatively small number of coin tosses or steps, the binomial distribution, a discrete probability distribution, characterizes the probability of finding the walker at each vertex. In the limit of a large number of steps, a Gaussian distribution provides  the vertex-wise probabilities.
	
	\par We mention here that a cyclic graph is one which is like a necklace with the beads representing vertices and the strings between the beads, the edges. In other words, a cyclic graph consists of a single cycle. We note that the line and the cyclic graphs are both connected as well as two-regular, assuming the line to be infinite. 
	A ``bipartite" graph consists of a graph in which the vertices can be colored with two different colors, and where each edge connects vertices of different colors. In a non-bipartite graph, such a coloring scheme is not possible.
	For a cyclic graph with an even number of sites ($=N_1$) - a bipartite graph - the probability for the walker to be at the $(2i)^{th}$ site is $2/N_1$, and the same for the  $(2i+1)^{th}$ site is vanishing, for $i=0,1,...N_{1}-1$, after a large even number of steps. The roles of the even and odd sites get reversed if the number of steps is a large odd number. For a cyclic graph with odd number of sites ($=N_2$) - a non-bipartite graph - in the limit of large number of steps, we get a uniform probability distribution with all the sites having a probability of $1/N_2$. This is the distinction between CRWs on a bipartite and a non-bipartite cyclic graph. 
	
	\par Let us now discuss the action of the CRW on a cyclic graph in a bit more detail. It consists of a classical walker moving on sites $x$ $\in$ $\{0,1,...N-1\}$, based on the outcome of a coin toss. The jump is only to the adjacent vertices. Here the random variable is the position of the walker, whose values are determined by the values of another random variable, the coin toss. We associate a ``probability vector" corresponding to the site-wise probability distribution at every step. Now we look at the discrete-step evolution of the probability vector facilitated by a transition matrix (stochastic matrix). For a cyclic graph with $N$ sites $\in$ $V$, the probability vector, $\vec{P}(n)$, will have $N$ elements. Here, \(n\) denotes the number of steps. Let $\vec{P}(n)=(P_{0}(n) ,  P_{1}(n) ... P_{N-1}(n))^{T}$ $\in$ $\mathbb{R}^N$.
	The elements of the transition matrix are given by
	\begin{equation}
	T_{ij} =
	\begin{cases}
	\frac{1}{2} & \text{if $j$ is the nearest neighbor of $i$} \\
	0 & \text{otherwise},
	\end{cases}
	\end{equation}
	where $i$, $j$ $\in$ $\{0,1...N-1\}$. From a given vertex $i$, the walker randomly chooses to move along either of the two edges to reach the nearest neighbors $j$ with a uniform probability of $1/2$. This is because the degree of every vertex of a cyclic graph is two and we are assuming an unbiased coin. Note that $T_{ij}=T_{ji}$. $\vec{P}(n)$ is obtained by the action of the transition matrix $T$ $:\mathbb{R}^N\rightarrow\mathbb{R}^N$ on the probability vector after $n-1$ steps, $\vec{P}(n-1)$ :
	\begin{equation}\label{lambda}
	\vec{P}(n)=T \vec{P}(n-1).
	\end{equation}
	The elements of $\vec{P}(n)$ are given by
	\begin{equation}
	P_{i}(n)=\sum_{j=0}^{N-1}T_{ij}P_{j}(n-1).
	\end{equation}
	

	\section{\label{sec:levell4}Interference in quantum random walk on a cyclic graph}
	 \par In this section, we discuss about QRWs in general and then move on to its features on a cyclic graph. We investigate the cases of symmetric and asymmetric initial coin states for our analysis. The associated localization is studied in some detail.
	
	\par An initially localized wave packet of the walker evolves as per the assigned local transition rules from a given vertex through either of the two edges to the corresponding two neighboring vertices based on the outcome of a coin toss at every discrete step. This kind of discrete evolution is facilitated by local action at the respective vertices and modeled by a unitary operator without the conventional Hamiltonian. We are interested in studying the global properties of the walk, such as the spread, as quantified, e.g., by standard deviation or Shannon entropy in position space. 
	In this paper, we will use the latter quantity to measure the 
spread. It is defined as \(-\sum_{i}p_i\log_2p_i\), for a probability distribution \(\{p_i\}\). We use logarithms of base 2 in all calculations of entropy, so that they are measured in bits. 
	For a discrete-time quantum walk (DTQW) on a line, the vertex set $V$ forms the position basis $\{\ket{x}$: $x$ $\in$ $\mathbb{Z}$$\}$ that spans the position Hilbert space, $\mathcal{H}_p$. The two basis states, $\ket{0}_{c}$ and $\ket{1}_{c}$ of the coin, which label the two directed edges at any given vertex, span the coin Hilbert space $\mathcal{H}_c$. The unitary evolution of the walk at every step is performed through a coin operation acting on the coin basis states followed by a conditional shift operation on the position basis states - conditioned on the states of the coin. The coin operation is parametrized by a ``coin parameter" $\theta$. We focus on the DTQW on a cyclic graph for our study. We will henceforth be denoting DTQW as QRW.
	
	\par For a QRW on a cyclic graph, the dimension of the position Hilbert space, $\mathcal{H}_p$, is fixed to $N$, the total number of sites (vertices), and $\mathcal{H}_p=$ $\{\ket{x}:$  $x$ $\in$ $\mathbb{Z}$ $\cap$ $[0, N-1]$$\}$. Here, the basis states of the coin instruct whether the next step of the walker will be in the clockwise or anti-clockwise direction. On a line, a significant portion of the quantum wave function - ``significant" in terms of the corresponding position probabilities - moves away from the point of the initial position of the walker, and so, although there is in principle room for interference between the different parts of the significant portion, it does not happen in a substantial way. The situation is completely different in the case of the QRW on a cyclic graph, where the different parts of the significant portion are forced to ``collide" (interfere) with each other due to the topology of the graph, provided \(N\) is about the same order as \(n\) or smaller than that, where \(n\) is the number of steps. This is one of the reasons that makes the cyclic graph an interesting class of graphs on which to study QRWs. For a cyclic graph with an odd number of sites, $N_2$, in the limit of a large number of steps, the time-averaged probability distribution becomes a uniform distribution with a uniform probability of $1/N_2$. In the case of an even number of sites, the time-averaged probability distribution is not uniform. This is an interesting distinction between a bipartite and a non-bipartite cyclic graph in a QRW.
	
\par We now study the QRW on a cycle in more detail, and discuss the corresponding localization effects. To begin, the ``coin operator", $\hat{C}_{\theta} : \mathcal{H}_c\rightarrow\mathcal{H}_c$, is given by
\begin{eqnarray}
\hat{C}_{\theta}&=&\cos\theta(\ket{0}\bra{0})_c+\sin\theta(\ket{0}\bra{1})_c\nonumber\\
&+&\sin\theta(\ket{1}\bra{0})_c-\cos\theta(\ket{1}\bra{1})_c,
\end{eqnarray}
with $\theta$ $\in$ $(0,\frac{\pi}{2})$.
We also consider the conditional shift operator, $\hat{S}_{x}:\mathcal{H}_c\otimes\mathcal{H}_p\rightarrow\mathcal{H}_c\otimes\mathcal{H}_p$, which shifts the position of the walker by a signed (i.e., directed) step length $\Delta x= +1$ if the coin is in the state $\ket{0}_c$, and $\Delta x= -1$ if the same is in $\ket{1}_c$. More precisely,
\begin{eqnarray}
\hat{S}_{x}&=&\sum_{x=0}^{N-1}[\ket{0}_c\bra{0}_c\otimes\ket{x+1\;(\bmod\; N)}_p\bra{x}_p\nonumber\\
&+&\ket{1}_c\bra{1}_c\otimes\ket{x-1\;(\bmod\; N)}_p\bra{x}_p]
\end{eqnarray}
The state after $n$ steps, $\ket{\psi(n)}$, is obtained by the action of the unitary operator $\hat{U}(\theta)=\hat{S}_{x}\cdot(\hat{C}_{\theta}\otimes I):\mathcal{H}_c\otimes\mathcal{H}_p\rightarrow\mathcal{H}_c\otimes\mathcal{H}_p$ on the state after $n-1$ steps, $\ket{\psi(n-1)}$ :
\begin{equation}
\ket{\psi(n)}=\hat{U}(\theta)\ket{\psi(n-1)}.
\end{equation}
The classical version of this equation is given in Eq. \ref{lambda}, where the unitary is replaced by a transition matrix.
Due to the action of the unitary, the position and coin states become entangled already after the first step of the walk, and the general state of the walker after $n$ steps, for fixed number of sites $N$, takes the form
\begin{equation}\label{gamma}
\ket{\psi(n)}=\sum_{x=0}^{N-1}[(a_{x}(n)\ket{0}_c+b_{x}(n)\ket{1}_c)\otimes\ket{x}_p],
\end{equation}
where $a_{x}(n)$ , $b_{x}(n)$ are the probability amplitudes corresponding to the clockwise and anti-clockwise directions.
From Eq. (\ref{gamma}), the probability distribution over the sites $x$ after $n$ steps of the walk is given by
\begin{equation}
P(x,n)=\lvert a_x(n) \rvert^2 +\lvert b_x(n) \rvert^2.
\end{equation}

	\begin{figure}[H]
	\centering
	\includegraphics[scale=0.8]{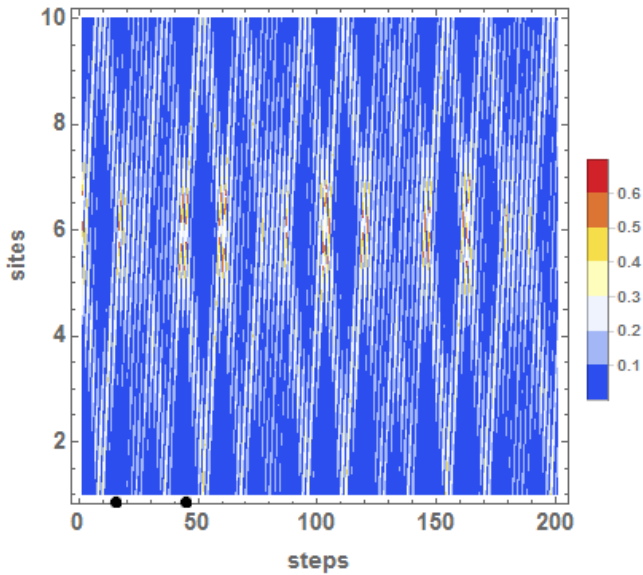}
	\caption{Projection plot of the position probability distribution of a Hadamard walk on a cycle with 10 sites. The discrete probability distribution is plotted as a function of sites and steps. A symmetric initial state of the coin qubit is chosen. The components of the probability distribution, namely, the clockwise and the anti-clockwise components ``collide" after every few sites. See text for precise definitions. The first few ``collisions" are marked on the horizontal axis with black dots. The regions in shades of red indicate that the probability of finding the walker is high. The vertical axis represents the site numbers, while the horizontal axis gives the number of steps, and the applicate represents the probabilities. The abscissa, ordinate, and applicate are dimensionless.}
	\label{fig:alpha2}
\end{figure} 

\subsection{\label{sec:subsecc1}Hadamard walk on a cyclic graph}

 The ``Hadamard" walk on a cyclic graph is the case when the coin parameter $\theta=\frac{\pi}{4}$. The Hadamard coin operator $\hat{H}$ is given in matrix form as
\begin{equation}
H= \frac{1}{\sqrt{2}}\begin{bmatrix}
1      & 1  \\
1      & -1 
\end{bmatrix}
\end{equation}
in the $\{\ket{0}_c,\ket{1}_c\}$ basis.\\

\paragraph{\label{para-for-symmetric}Symmetric initial coin state, enhanced interference and the ensuing localization:}


 We choose a symmetric initial state for the coin, viz. \(\frac{\ket{0}_c\pm i\ket{1}_c}{\sqrt{2}}\). The sign in front of $i$ is unimportant for further calculations. We begin the discussion  for $N=10$ and $\theta=\frac{\pi}{4}$, for different numbers of steps. The joint initial state is taken to be $\ket{\psi(0)}=\frac{\ket{0}_c\pm i\ket{1}_c}{\sqrt{2}}\otimes\ket{6}_p$. The choice of the initial state of the walker is of course arbitrary. The position probability at the site $6$ begins with unity at the initial time, gradually diminishes with increase of the number of steps, reaches a minimum (a local minimum at site $6$ with respect to number of steps) and then again increases to reach a maximum (a local maximum at site $6$ with respect to number of steps) before diminishing once more. Each such maximum is called a ``meeting point", arguably of the clockwise and anti-clockwise components of the joint state. The corresponding number of steps are denoted by $n_{meet}$.
 Fig. \ref{fig:alpha2} exhibits the symmetric site-wise probability distribution for $N=10$ and $\theta=\frac{\pi}{4}$, for different numbers of steps.
  The first few meeting points are marked in black dots on the horizontal axis in the figure. The first meeting happens after 16 steps of the walk, the second meeting after 42 steps, and so on. Due to the oscillatory nature (with respect to number of steps) of the entropy of the probability distribution $P(x,n)$ for fixed $n$, we take the average entropy, $H^{Q}_{meet}$ up to every $n_{meet}$  to measure the fluctuation in the probability distribution, instead of considering the entropy itself:
\begin{equation}
H^{Q}_{meet} (n_{meet}) = \frac{1}{[n_{meet}]}\sum_n\sum_{x=0}^{N-1}(-P(x,n)\log_2 P(x,n)),
\end{equation}
where the sum over $n$ runs up to $n_{meet}$ from just after the previous meeting point, and where $[n_{meet}]$ denotes the number of steps in that interval between the meeting points.

\begin{figure}[H]
	\centering
	\includegraphics[scale=0.75]{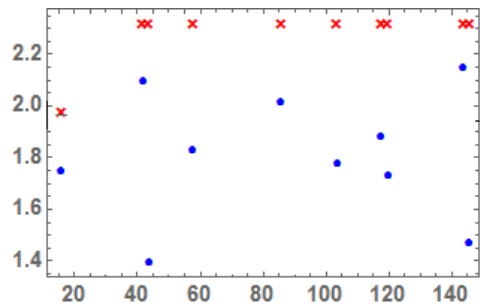}
	\caption{Localization in quantum random walk in comparison to classical one, as inferred from step-wise entropy, averaged over steps between consecutive meeting points. From Fig. \ref{fig:alpha2}., the meeting points, $n_{meet}$, are noted down. The average site-wise entropies between two meeting points, as denoted by $H^{Q}_{meet}(n_{meet})$, are plotted on the vertical axis against $n_{meet}$ on the horizontal axis. These are represented as blue dots in the figure. The classical case is represented as red crosses. The horizontal axis is dimensionless, while the vertical one is in bits.}
	\label{fig:alpha8}
\end{figure}

\begin{figure}[H]
	\centering
	\includegraphics[scale=0.56]{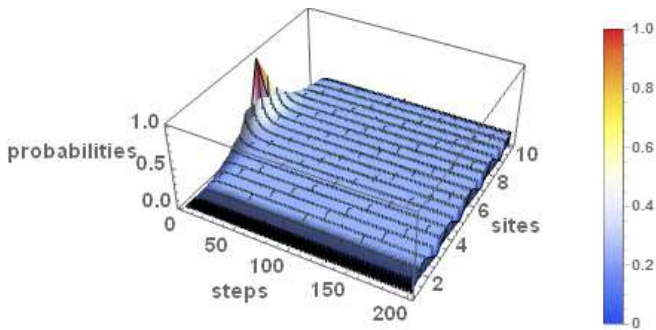}
	\caption{Position probability distribution of a classical random walk on a cycle with 10 sites and with an unbiased classical coin. This is the classical counterpart of the quantum case in Fig. \ref{fig:alpha2}. All other considerations are as in Fig. \ref{fig:alpha2}.}
	\label{fig:alpha50}
\end{figure}

The plot of $H^{Q}_{meet}$ against $n_{meet}$, in Fig. \ref{fig:alpha8}, 
captures an important feature of the interference-induced dynamics in a QRW on a clean cyclic graph.
The plot is done up to a sufficiently large number of meeting points so that no further appreciable change occurs in the value of $H^{Q}_{meet}$. The plot of the same quantity in the classical case is denoted by  $H^{C}_{meet}$ and depicted by red crosses in Fig. \ref{fig:alpha8}. The associated probability distribution of the CRW is shown in Fig. \ref{fig:alpha50}. $H^{C}_{meet}$ exhibits very little fluctuations with the number of steps and saturates approximately around the value of 2.32 bits. It may be noted here that the classical walker does not encounter a ``collision" between  the anti-clockwisely and clockwisely traversing components. The corresponding plot, for a QRW (blue dots in Fig. \ref{fig:alpha8}), displays significantly higher fluctuations (than in the classical case) and saturates approximately around the value of 2 bits, and being lower than the corresponding classical value of 2.32 bits, indicates a localization behavior.

\par
\begin{widetext}

\begin{figure}[H]
	\vspace{5mm}
	\centering
	\subfloat[]{
		\includegraphics[width=80mm,height=50mm]{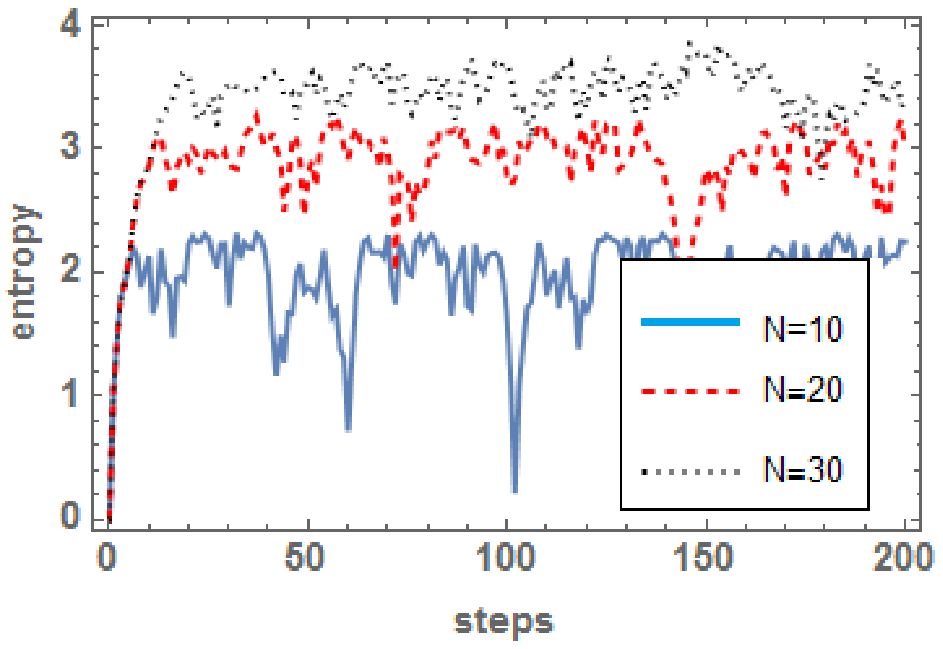}
	}
	\subfloat[]{
		\includegraphics[width=80mm,height=50mm]{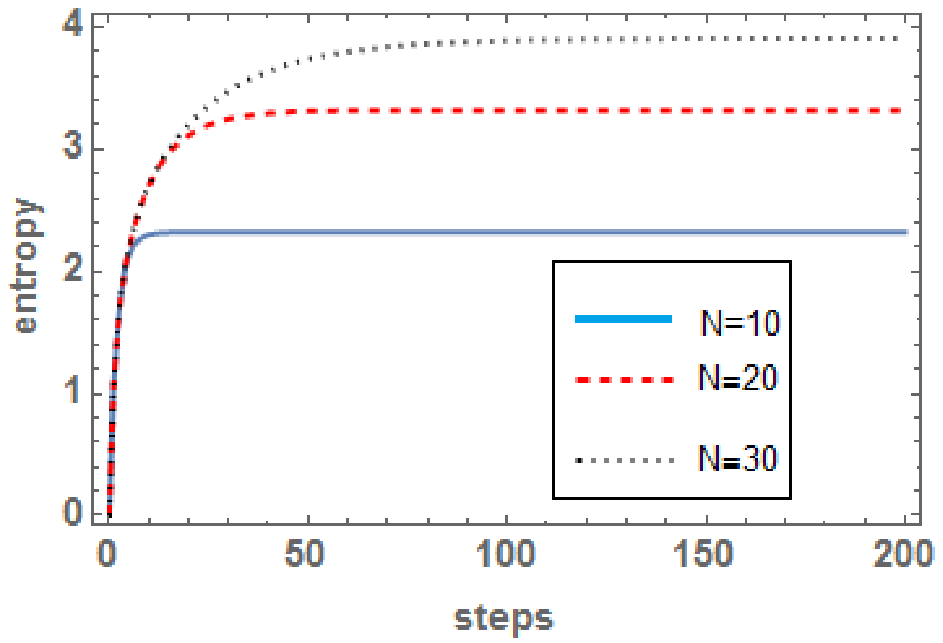}
	}
	\caption{Comparison between classical and quantum random walks with respect to their variation of step-wise entropy of position probability distribution. The quantum case is presented in panel $(a)$, while panel $(b)$ depicts the classical case. For a given system size $N$, we plot the entropy of $\{P(x,n)\}_{x}$ for every $n$ up to a sufficiently large number of steps in panel $(a)$. The corresponding classical case is plotted in panel $(b)$. The maximal number of steps is so chosen that no further appreciable change in entropy occurs for higher number of steps. The plot in the classical case does not fluctuate after having reached its steady value. There are however significant fluctuations in the quantum case. These fluctuations in the quantum case gets diminished for larger $N$ (see Fig. \ref{fig:alpha5} (a)). For any $N$, the steady-state value in the classical case is higher than the average (over steps) in the quantum walk.}
	\label{fig:alpha4}
\end{figure}
\end{widetext}

\begin{widetext}

\begin{figure}[H]
	\vspace{5mm}
	\centering
	\subfloat[]{
		\includegraphics[width=80mm,height=50mm]{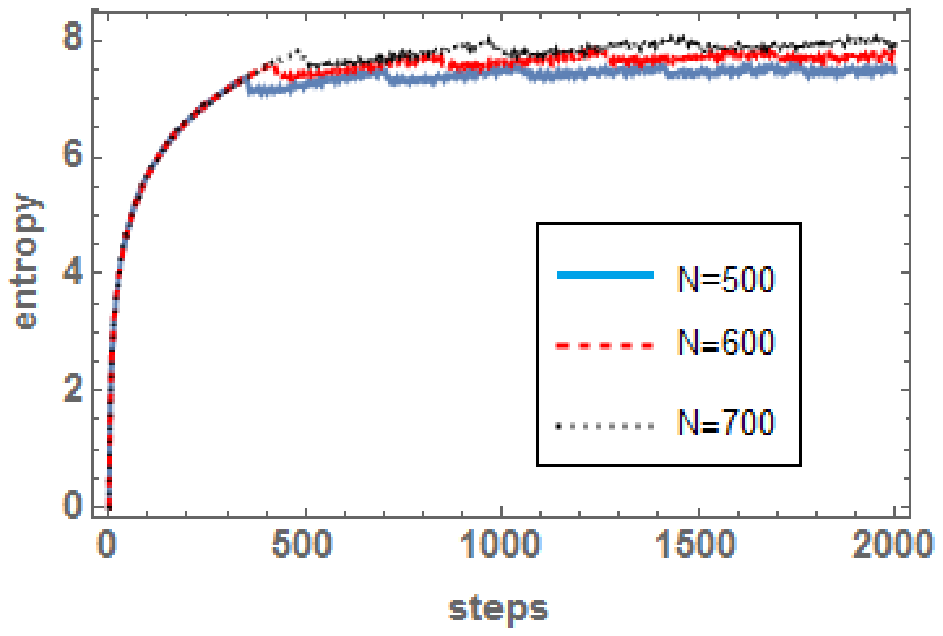}
	}
	\subfloat[]{
		\includegraphics[width=80mm,height=50mm]{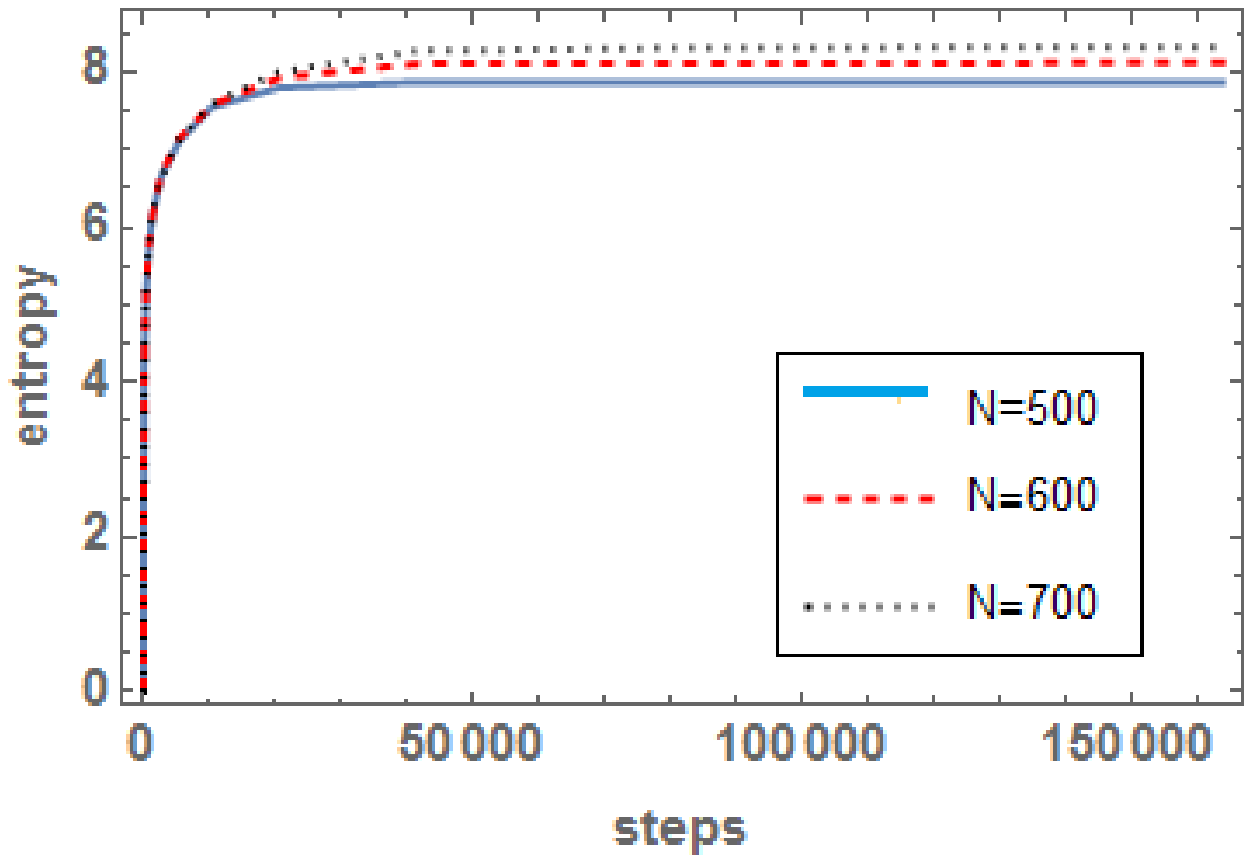}
	}
	\caption{The description is the same as in Fig. \ref{fig:alpha4} except that cyclic graphs of site numbers $N=500$ (in blue), $600$ (in red), and $700$ (in black) are considered. Just like in Fig. \ref{fig:alpha4}, panels \((a)\) and \((b)\) are respectively for the quantum and classical cases. In panel $(b)$, we needed to consider a larger number of steps to obtain convergence of the entropy with respect to number of steps.}
	\label{fig:alpha5}
\end{figure}

\end{widetext}

 We now compute the entropies themselves (instead of the average entropies),
in the classical and the quantum cases, of the position probability distributions, for every $n$ up to a value of $n$ after which we do not envisage any further significant change of behavior, for a fixed number of sites \(N\). We perform the analysis on cyclic graphs for $N=10$, $20$, $30$, $500$, $600$, and $700$ (see Figs. \ref{fig:alpha4} and \ref{fig:alpha5}). 
 We refer to these entropies as ``site-wise" entropies to differentiate them from the ``average entropies", $H^{Q}_{meet}$. We find that irrespective and in spite of the fluctuations present, the entropy with respect to number of steps for the QRW saturates to a value that is lower than the corresponding value in the classical case, for cyclic graphs of various sizes, indicating a certain amount of localization in the quantum case. We  also search for the asymptotic behavior of the Shannon entropy by calculating its value for  cyclic graphs    of 
 size $N$ as high as 
 1000, and observe  that the entropy on a cyclic graph of a given size for the QRW still saturates to the same value as inferred from Figs. \ref{fig:alpha4} (a) and \ref{fig:alpha5} (a). We notice that as we increase the number of sites, the fluctuations in the entropy with number of steps in the quantum case is reduced. 

\begin{figure}[H]
	\centering
	\includegraphics[scale=0.6]{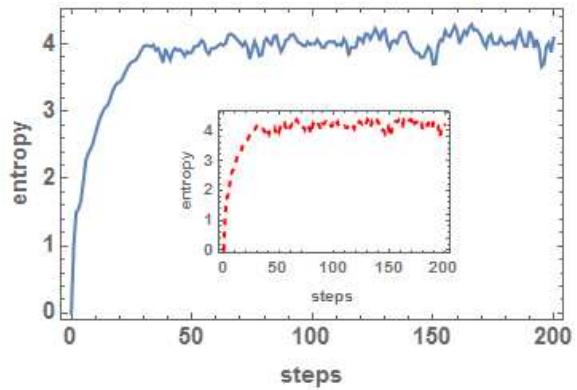}
	\caption{Variation of site-wise Shannon entropy with number of steps for a Hadamard walk on a cycle with 50 sites for an asymmetric initial coin state. The variation is seen up to 200 steps. The chosen initial state, in the main figure, is $\ket{\psi(0)}=\ket{0}_c\ket{25}_p$. The overall behavior is very similar to that of the corresponding case with a symmetric initial coin state. The behavior remains similar in the inset, where the initial state of the coin-walker joint system is chosen to be \((\cos\Theta |0\rangle_c + e^{i\Phi}\sin\Theta |1\rangle_c) \ket{25}_p\) with \(\Theta = \pi/3\), \(\Phi = \pi/2\). This shows that the symmetry of the initial coin state has little to no effect on the localization behavior. The vertical axes represent the Shannon entropy (in bits) and the horizontal axes represent the number of steps (dimensionless).}
	\label{fig:alpha21}
\end{figure}

\paragraph{The case of asymmetric initial coin states:}
The study performed until now was based on a symmetric initial state of the coin. Let us now check whether the trends of the probability distribution of QRW on a cyclic graph are independent of this choice. Towards this aim, 
 we now choose an asymmetric initial coin state, viz. \(\cos(\Theta/2)|0\rangle_c + e^{i\Phi}\sin(\Theta/2) |1\rangle_c\).
As an example, let us begin with the case when 
 %
 the joint initial state of the coin-walker system is $\ket{\psi(0)}=\ket{0}_c\ket{25}_p$ on a cyclic graph with 50 sites. Fig. \ref{fig:alpha21} shows the variation of site-wise entropy with respect to number of steps for this case. The inset of the same figure provides the behavior for a different pair of \(\Theta\) and \(\Phi\). Comparing Fig. \ref{fig:alpha21} with Figs. \ref{fig:alpha4} and \ref{fig:alpha5}, we conclude
 that the global effect still remains the same, i.e., the quantum walker still shows the localization behavior. We have checked that broadly the same behavior is obtained for other values of \(\Theta\) and \(\Phi\) also. This shows that regardless of whether  the coin state is symmetric,
 localization of the wavepacket occurs. This implies, once again, that localization in quantum random walks on the cyclic graph is a consequence of the interference (``collision") between the clockwise and anti-clockwise components of the coin-walker quantum wavefunction.

\subsection{\label{sec:subsecc}Non-Hadamard walks}
\begin{figure}[H]
	\centering
	\subfloat[]{
		\includegraphics[width=40mm]{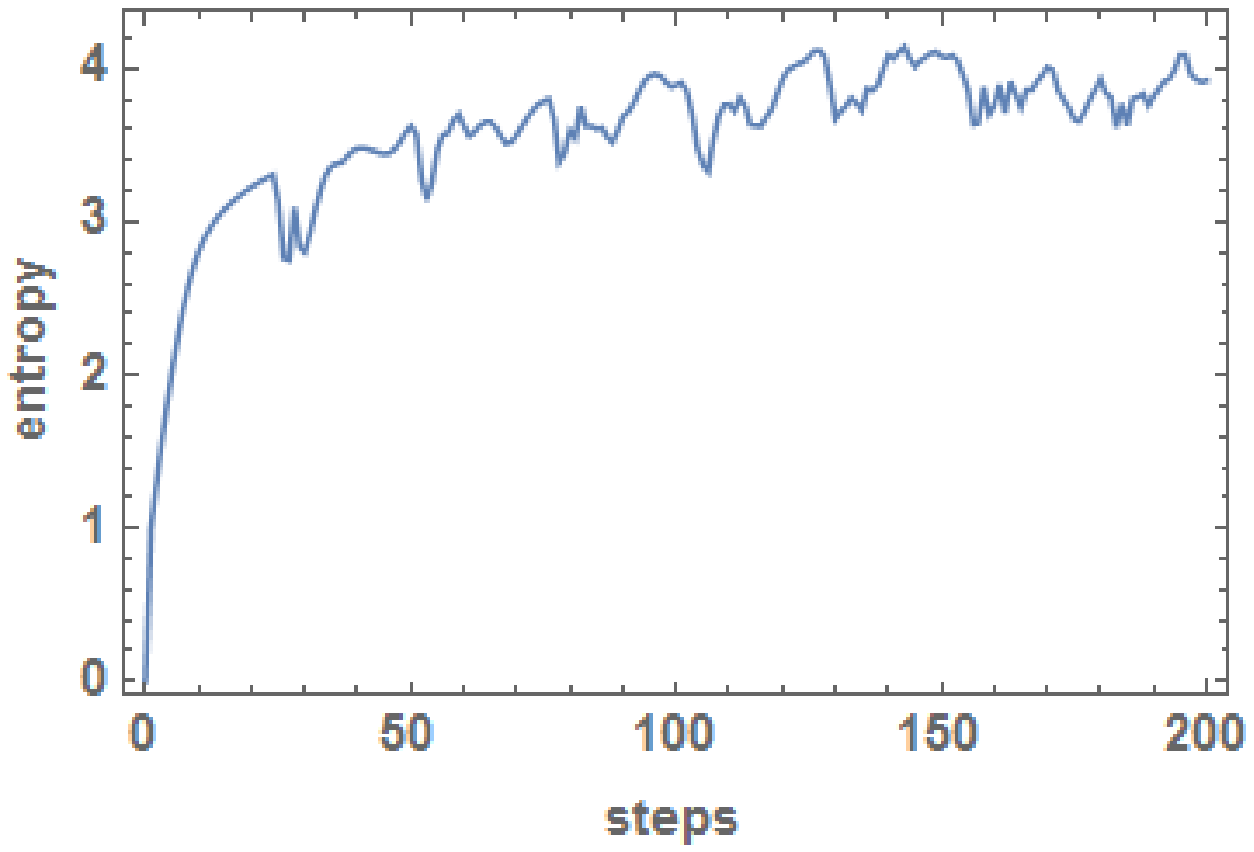}
	}
	\subfloat[]{
		\includegraphics[width=40mm]{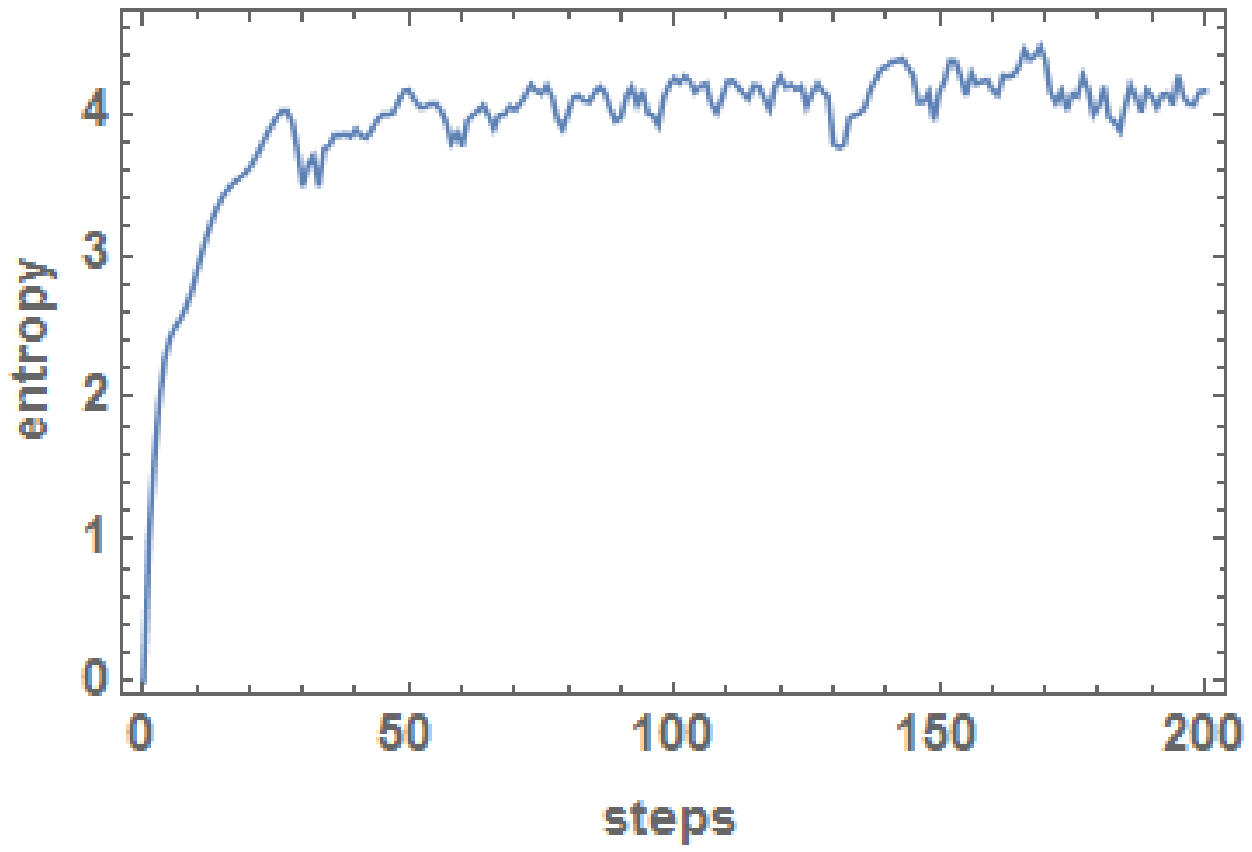}
	}
	\hspace{0mm}
	\subfloat[]{
		\includegraphics[width=40mm]{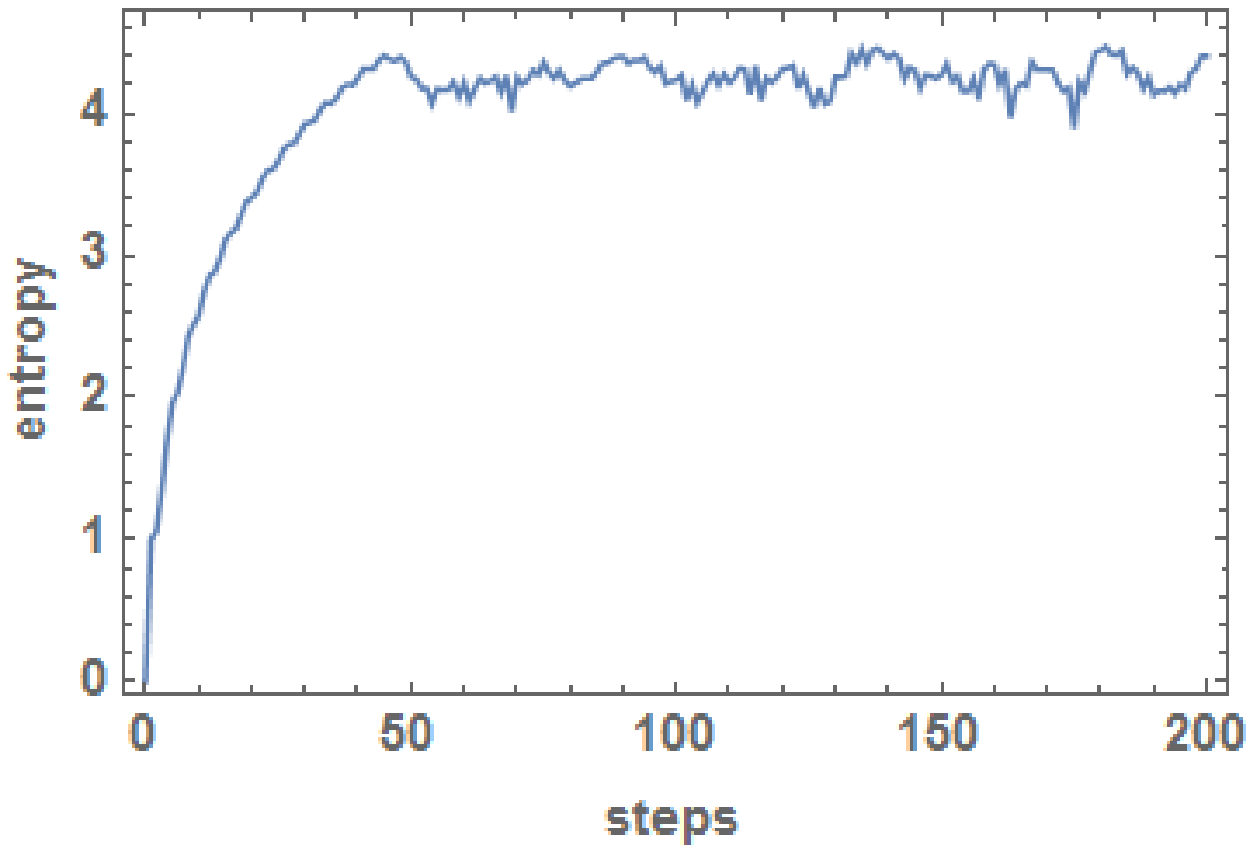}
	}
	\subfloat[]{
		\includegraphics[width=40mm]{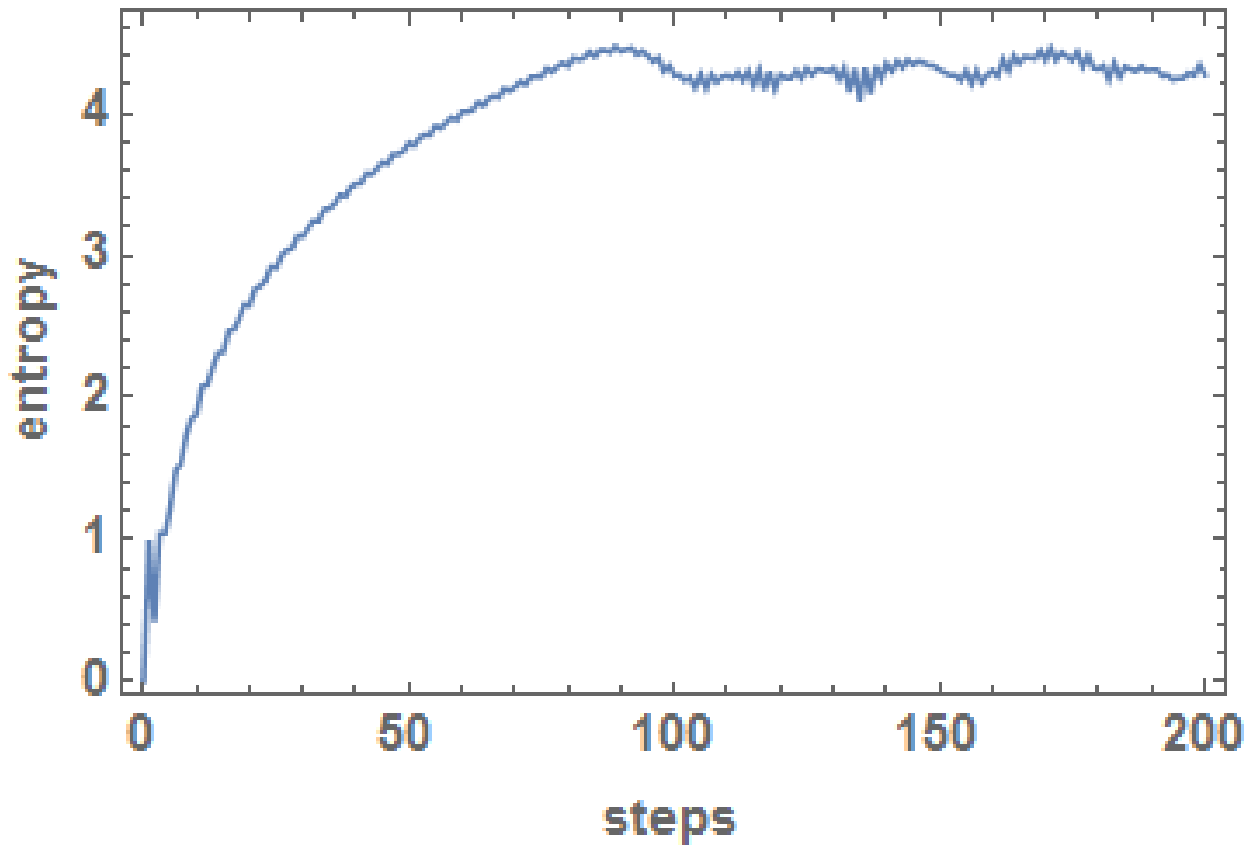}
	}
	\caption{Variation of site-wise Shannon entropy with the number of steps for non-Hadamard walks on a cycle up to 200 steps. We set the number of sites on the cycle as 50 and notice that for four different values of the coin parameter namely, (a) $\theta =\frac{\pi}{12}$, (b) $\frac{\pi}{6}$, (c) $\frac{\pi}{3}$, and (d) $\frac{5\pi}{12}$, the entropies converge to almost a fixed value. Therefore, even for non-Hadamard walks, the localization behavior persists and the overall behavior is roughly the same as for Hadamard walks. Note that on increasing the value of $\theta$ from $\frac{\pi}{12}$ to $\frac{5\pi}{12}$, the fluctuations seen in the entropy keeps decreasing. The vertical axis represents the Shannon entropy (in bits) and the horizontal axis represents the number of steps (dimensionless).}
	\label{fig:alpha7}
\end{figure}

Let us now examine 
the variation of site-wise entropy with number of steps for four non-Hadamard walks on a cyclic graph with 50 sites, namely for $\theta=\frac{\pi}{12}$, $\frac{\pi}{6}$, $\frac{\pi}{3}$ and $\frac{5\pi}{12}$. The Hadamard walk is for $\theta=\frac{\pi}{4}$, and was the object of discussion in the preceding subsection. The aim of such investigation is to identify the effect of \(\theta\) (in \(\hat{C}_\theta\)) on the overall behavior of the site-wise entropies with increase in number of steps. The initial state of the coin is chosen to be \(\frac{1}{\sqrt{2}}(|0\rangle \pm i|1\rangle)\). As depicted in Fig. \ref{fig:alpha7}, we find that the site-wise entropy still saturates, approximately around the value of 4 bits, with the number of steps, like in the Hadamard case. This shows that using a biased coin operation at every step of the quantum walk has no bearing on the overall average variation of site-wise entropy. However, we observe that  for lower values of $\theta$, the fluctuation in the entropy with number of steps is relatively more than that for higher values.

\section{\label{sec:levell5}Behavior with respect to system-size}
 \par Until now, we have mainly been looking at the behavior of site-wise entropy as a function of the number of steps for a given number of sites. We now
 do a role reversal
 and study the patterns of site-wise entropy with respect to system size for quantum and classical random walks on cyclic graphs. After fitting suitable functions to the plots, we find out the limiting behavior of the ratio of the two functions as the system size grows to infinity. We find this ratio for three different functions in the quantum case since all of them have small and comparable least square errors in their respective fits.

\begin{figure}[H]
		\centering
	\includegraphics[scale=0.6]{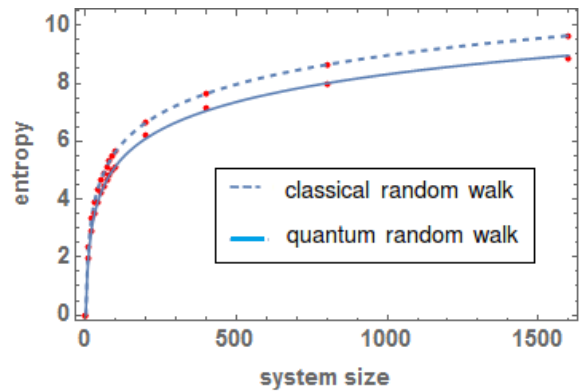}
	\caption{Variations of site-wise Shannon entropy, averaged over steps in the ``saturation" region, with respect to system size for quantum and classical walks on a cyclic graph. The fitted curve in the quantum case is given by Eq. (\ref{alpha}) for $\nu=2$. For the classical walker, Eq. \ref{beta} is used. The vertical axis represents Shannon entropy (in bits) and the horizontal axis represents system size (dimensionless).}
	\label{fig:alpha22}
\end{figure}

	\par \textit{Site-wise entropy with respect to system size for a QRW on a cyclic graph:} We have already observed that in general, the site-wise entropy for a fixed number of sites registers a steep monotonic increase up to a certain number of steps, after which saturation to a certain extent occurs. See Figs. \ref{fig:alpha4} (a) and \ref{fig:alpha5} (a). Due to the fact that significant fluctuations may be present even in this ``saturated" regime, a characteristic ``converged" value can be obtained, for a given system-size (number of sites), only by performing an average, over steps, in this saturated regime. We present these converged values in Fig. \ref{fig:alpha22} against system size for a quantum random walk on a cyclic graph, with the initial state of the coin being again chosen to be \(\frac{1}{\sqrt{2}}(|0\rangle \pm i|1\rangle)\). For a system size of $N$, we denote the converged site-wise Shannon entropy by $H^{Q}(N)$. We fit the function
	\begin{equation}\label{alpha}
	\alpha\log_2(1+\beta N^{\nu}),
	\end{equation}
	for $\nu=1$, $2$, $3$ to the obtained values of $H^{Q}(N)$ against $N$. For each $\nu$, we use the method of least squares to find $\alpha$ and $\beta$. We find that $\alpha=0.6810$, $\beta=0.3583$ for $\nu=1$, $\alpha=0.3305$, $\beta=0.1765$ for $\nu=2$, and $\alpha=0.2198$, $\beta=0.0767$ for $\nu=3$. The respective least squares errors are $0.0574$, $0.0352$ and $0.0338$.  
	\par \textit{Site-wise entropy with respect to system size for a CRW on a cyclic graph:} We perform a parallel set of calculations for the classical walker. Again, the site-wise entropy for a given system size has a steep monotonic increase up to a certain number of steps, after which it saturates. Unlike in the quantum case, there are no fluctuations in the saturated region. See Figs. \ref{fig:alpha4} (b) and \ref{fig:alpha5} (b). The corresponding saturated values, $H^{C}(N)$, behaves as
	\begin{equation}\label{beta}
	H^C(N)=\log_2 \Big(\frac{N}{2}\Big).
	\end{equation}
	
	\par \textit{Quantum vs. classical in the large system-size limit:} The ratio $H^C(N)/H^Q(N)$ in the large $N$ limit is \(1/(\alpha \nu)\), so that it is $1.4684$, $1.5129$, and $1.5165$, respectively for $\nu=1$, $2$, and $3$. The ratio of the entropies in classical to the quantum case is greater than one, which indicates that the CRW spreads out more compared to the QRW for a given system size. This characterizes the spreading behavior in both the scenarios implying a slowdown or ``localization" in the quantum case. We also note that the two curves corresponding, respectively, to the classical and quantum walkers, never meet for any (non-zero) value of $N$.
	
	\section{\label{sec:levell3}Distance between the spreads of classical and quantum walkers}

\par In this section, we compare the spread of the probability distribution of the walker's position in a QRW with that of a CRW on a cyclic graph using the $l^1$- norm.

\begin{figure}[H]
	\centering
	\includegraphics[scale=0.5]{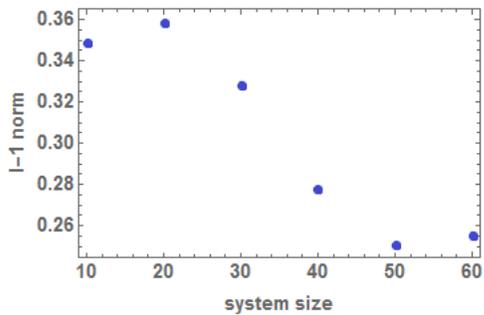}
	\caption{Distance between the spreads of classical and quantum walkers. We plot here the variation of the $l^1$- norm between the spreads as we increase the number of sites on a cyclic graph up to 60. The two populations namely, the time series data of entropies of a QRW and a CRW on a cyclic graph with a given number of sites are generated. The number of steps chosen while evaluating the distance is 200. The vertical axis represents the $l^1$- norm (in bits) and the horizontal axis represents the system size (dimensionless).}
	\label{fig:alpha23}
\end{figure}

\par  We start by choosing the symmetric initial states, $\ket{\psi(0)}=\frac{\ket{0}_c\pm i\ket{1}_c}{\sqrt{2}}\otimes\ket{0}_p$ for the quantum walker. Our aim is to use a measure that quantitatively captures the difference between the spreads of a QRW and a CRW on a cyclic graph. We use the $l^1$-norm for comparing the variation of Shannon entropies with respect to the number of steps. For a given system size, the site-wise Shannon entropies for increasing number of steps is similar to a time series data, where the steps represent time. We consider the time series data for the classical and the quantum cases. The distance between two time series data sets, $a=\{a_{i}\}$ and $b=\{b_{i}\}$, each of length $\mathbb{N}$, can be quantified, by using the $l^1$-norm, as
\begin{equation}
D_{l^1}(a,b)=\frac{1}{\mathbb{N}}\sum_i \lvert a_i - b_i \rvert.
\end{equation}
Fig. \ref{fig:alpha23} depicts the behavior of the distance, based on the $l^1$-norm, between the time series data sets for the classical and quantum cases, with respect to the number of sites on a cyclic graph. It is worthwhile to note that the distance measure satisfies the usual properties of being a metric \cite{citation-for-metric}. 



\section{\label{sec:level2}Discussion}
To conclude, 
we investigated the spreading behavior of a quantum random walker and compared the situation with that for a classical one, on a cyclic graph. The walk of a quantum entity dictated by a quantum coin is distinct from that of a classical one commanded by a classical coin, due to the superposition of different positions of the quantum walker and due to entanglement between the quantum walker and the quantum coin during the evolution. For the quantum walker on the infinite line, e.g. when using a symmetric coin, the position probability distribution dissociates into a bi-modal distribution, and the spread of the position is qualitatively higher than the corresponding classical walker on the same graph. Replacing the infinite line by a cyclic graph, there is a role reversal between the quantum and classical walkers with respect to their spreads. This shift in behavior of the quantum walker for a switching of the lattice is potentially attributable to an additional interference in the quantum walker wavefunction. For the initial few steps, the quantum walkers on the infinite line and the cyclic graph are no different. The situation changes when the number of steps is large enough, and we see that the initial bi-modal position probability distributions have the possibility to ``collide" in the case of a cyclic graph. A single classical walker on any graph does not have this option. A quantum walker on an infinite line, in principle, has this option, but the dynamics of the system does not let this happen. A quantum walker on a cyclic graph, however, is made to collide by the very dynamics due to the topology of the graph. The classical random walker keeps spreading on a cycle and reaches a steady state where the time-averaged probability distribution becomes a uniform distribution over all the sites of the cyclic graph. The quantum walker on a cyclic graph however has a lesser spread, and we refer to this as an instance of ``localization". The spreads are captured by the Shannon entropies of the position probability distributions of the walkers. While we began with the case of the quantum coin operator being so chosen that the corresponding quantum walk is the ``Hadamard" walk, we have later on also altered the coin parameter (to consider non-Hadamard walks), which introduces different degrees of interference in the dynamics, and have analyzed the consequent nature of localization. We found that for all the 
different values of the coin parameter considered, localization persists albeit with varying degrees of fluctuations in the entropy with respect to the number of steps. We saw that the symmetry of the initial coin state has minimal effect on the spreading behavior but has significant effect on the symmetry of the probability distribution. Subsequently, we compared the quantum and classical walkers by considering the behaviors of their entropies with respect to system size. Finally, we have used the $l^1$-norm as a distance measure between the variations of entropies with respect to number of steps in the quantum and the classical scenarios.

\end{document}